\documentclass[twocolumn,preprintnumbers,amsmath,amssymb]{revtex4}
\usepackage{graphicx} 
\usepackage{setspace} 
\usepackage{dcolumn} 
\usepackage{bm} 
\usepackage{natbib}
\usepackage{hyperref}
\hypersetup{colorlinks=true,linkcolor=blue,citecolor=blue,urlcolor=blue}
\usepackage{epsfig}
\usepackage{amsmath}
\begin{document}
\title{Tailored Thermal and Mechanical Performance of Biodegradable PLA-P(VDF-TrFE) Polymer Blends}
\author{G Suresh$^a$\footnote{Email: suresh.explorer@gmail.com}}
\author{B. Satyanarayana$^b$\footnote{Email: satyanarayana\_b@gmail.com}}
\author{C. Thirmal$^a$\footnote{Email: thiruphysics02@gmail.com}}
\author{Kaushal Jagarlamudi$^b$}
\author{T Komala$^b$}
\author{Jimlee Patowary$^c$}
\author{Ashutosh Kumar$^c$}
\affiliation{$^a$Battery Technology Lab, Department of Physics, VNR-Vignana Jyothi Institute of Engineering and Technology, Hyderabad}
\affiliation{$^b$Department of Mechanical Engineering, VNR-Vignana Jyothi Institute of Engineering and Technology, Hyderabad}
\affiliation{$^c$Functional Materials Laboratory, Department of Materials Science and Metallurgical Engineering, Indian Institute of Technology Bhilai-491 002, India}
\date{\today}
\begin{abstract}
The development of polymer blends has emerged as a strategic approach for designing multifunctional materials with enhanced tailored characteristics. Current work investigates and reports for the first time, the structure–property relationships in free-standing blend films of poly(vinylidene fluoride-trifluoroethylene) (P(VDF-TrFE)) and polylactic acid (PLA), prepared to evaluate their suitability for functional applications. For this investigation, films of approximately 40 $\mu$m thick were fabricated by systematically varying the P(VDF-TrFE):PLA ratio. Thermal analysis revealed a higher PLA crystallinity at 25\% P(VDF-TrFE) content, while Fourier-transform infrared spectroscopy showed the electroactive $\beta$-phase fraction to be highest in the 50:50 composition. These findings correlated with tensile strength measurements and morphology, demonstrating that molecular ordering and phase distribution significantly influence the mechanical performance. The 25:75 blend exhibited superior mechanical strength due to enhanced PLA crystallization and polymer chain alignment. In contrast, the 50:50 blend achieved a balance between tensile modulus and electroactive phase development, marking it a promising candidate for sensors and 3D printing applications. At higher P(VDF-TrFE) content, reduced crystallinity in PLA resulted in softer, more compliant films which would be suitable for flexible electronic applications. These results establish a pathway to tune mechanical and functional properties in semicrystalline polymer blends through facile compositional control.\\                
\end{abstract}
\maketitle
\section{Introduction}
Polymer blends—mixtures of two or more distinct polymers—have emerged as a versatile and cost-effective approach for tailoring materials with enhanced or multifunctional properties\cite{R1}. These blends are typically classified as miscible (single-phase systems), immiscible (multi-phase systems with distinct domains), or compatible blends, which are immiscible blends engineered to exhibit uniform physical characteristics\cite{R2}.By leveraging the complementary attributes of individual polymers, such blends can be designed to improve thermal stability, mechanical resilience, chemical resistance, and processability—characteristics vital for applications ranging from biomedical implants and flexible electronics to 3D printing and smart packaging \cite{R3}.\\
Thin polymer films produced from such blends are widely utilized due to their tunable morphology, processability, and scalable fabrication routes such as extrusion, spin coating, casting, and solution deposition. Their thickness can range from a few nanometers to several millimeters depending on application demands, and their performance can be further elevated by incorporating fillers or copolymers \cite{R4,R5}. Modern strategies using nanocomposites, multilayered structures, and functional additives have shown that electrical, mechanical, or thermal properties can be precisely tailored to meet specific application needs—particularly in fields demanding biodegradability, flexibility, and functional responsiveness \cite{R6}.
In this context, blending polylactic acid (PLA), a biodegradable aliphatic polyester derived from renewable sources like corn starch and sugarcane \cite{R7,R8}, with poly(vinylidene fluoride-trifluoroethylene) (P(VDF-TrFE)), a semicrystalline copolymer known for its ferroelectric and piezoelectric properties\cite{R9,R10}, presents a promising pathway for developing multifunctional materials. PLA's ecological benefits and moderate mechanical performance make it attractive for biomedical and packaging applications, particularly additive manufacturing \cite{R11}. However, its poor heat resistance (T$_g$$\sim$60$^{\circ}$C), brittleness, and low impact strength limit its utility in robust or flexible systems \cite{R12}. To address these drawbacks, chemical crosslinking and blending techniques have been explored. For instance, triallyl isocyanurate (TAIC) and dicumyl peroxide (DCP) have been used to enhance PLA's thermal degradation resistance and modulus \cite{R8,R12}, and the addition of cross-linked natural rubber has dramatically improved impact strength \cite{R13}. The blend ratio also plays a crucial role in tuning crystalline and mechanical properties, as shown by Hamidi et al in PLA-TPU systems \cite{R7}.\\ 
On the other hand, P(VDF-TrFE) offers intrinsic electroactivity via direct crystallization into the $\beta$-phase, making it suitable for energy harvesting, sensors, and flexible electronics \cite{R14}. Processing conditions—particularly annealing between the Curie temperature and melting point—have been shown to modulate crystallinity, electromechanical response, and barrier properties\cite{R15}. Despite extensive work on P(VDF-TrFE) blends with elastomers and nanofillers, its integration with PLA remains underexplored. As per our knowledge, very little or no studies have explored the inclusive properties of PLA–P(VDF-TrFE) blends. Notably, systematic investigations on how these two polymers influence each other's morphology, crystallization behavior, and functional phase transitions are scarce. The thermal overlap between PLA’s melting/crystallization temperatures and P(VDF-TrFE)'s Curie transition hints at potential phase coupling effects, whereby crystallinity and electroactive phase development could be composition-dependent.\\
To the best of our knowledge, this study represents the first systematic exploration of how blend composition and phase interactions influence the properties of biodegradable PLA–P(VDF-TrFE) polymer films. Addressing a clear gap in existing literature, we investigate drop-cast films with carefully varied weight ratios of PLA and P(VDF-TrFE), aiming to uncover how structural and functional characteristics evolve with composition.\\
This research introduces a novel framework for tuning multifunctional polymer systems by correlating crystallinity, electroactive phase development, and mechanical performance. The core objective is to optimize interfacial compatibility, mechanical toughness, and ductility through targeted morphological tuning, phase dispersion, and control of thermal transitions. A comprehensive suite of characterization techniques—including differential scanning calorimetry (DSC), Fourier-transform infrared spectroscopy (FTIR), scanning electron microscopy (SEM), X-ray diffraction (XRD), and mechanical testing—were employed to evaluate how blend ratios influence the structure–property relationships of the resulting films. What sets this work apart is its integrated analysis of phase interaction effects within a biodegradable matrix, establishing meaningful links between crystallinity, electroactive response, and mechanical integrity. These findings not only open avenues for sustainable applications in 3D printing, wearable electronics, and soft robotics but also lay the groundwork for future research on electroactive–biopolymer integration. Ultimately, this contribution supports the development of next-generation polymer composites tailored for multifunctional and high-performance applications \cite{R16,R17,R18,R19}.\\
\section{Experimental Details}
\subsection{Materials and Methodologies}
\textbf{Materials:} The polymers used in this study are Poly(vinylidene fluoride–trifluoroethylene) (P(VDF-TrFE)) and Polylactic Acid (PLA). P(VDF-TrFE) supplied by Sigma-Aldrich (Solvene300/P300) powder was, chosen for its well-known ferroelectric, piezoelectric, and thermal stability characteristics. PLA, a biodegradable aliphatic polyester, is selected for its favourable mechanical strength, biocompatibility, and environmental degradability \cite{R8}. PLA was supplied by 2MB Tech (M. Wt. 46.07). The solvent used throughout the blend preparation was N,N-Dimethylformamide (DMF) of analytical grade, procured from Finar Chemicals Pvt. Ltd. Its high solvating power and relatively high boiling point ($\sim$153$^{\circ}$C) make it ideal for several applications ranging from dissolving semi-crystalline, high molecular weight polymers to the fabrication of metal nanoparticles \cite{R20,R21}.\\
\begin{figure*}
\centering
  \includegraphics[width=0.99\linewidth]{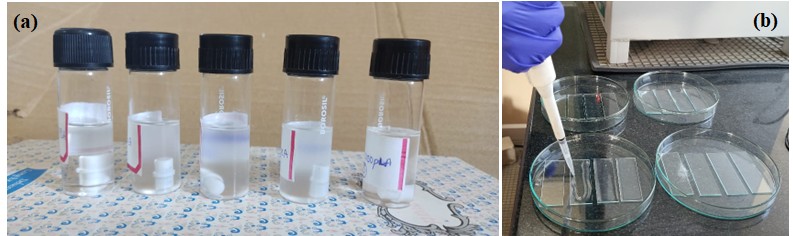}
  \caption{(a) Variation of P(VDF-TrFE):PLA blend composition, left bottle contains 0:100 (pure PLA), 25:75, 50:50, 75:25, and the right bottle contains 100:0 (pure P(VDF-TrFE)); (b) Casting technique adapted in film making.}
  \label{fig1}
\end{figure*}
\textbf{Preparation Methodology of Polymer Blends:} Blends were prepared by solution casting, a method that allows for precise control over polymer composition and film morphology. Both P(VDF-TrFE) and PLA were individually dissolved in DMF at room temperature under continuous magnetic stirring until visually homogeneous solutions were obtained. Each solution was prepared using a polymer-to-solvent ratio of 1 gram per 10 mL of DMF, ensuring consistent viscosity and solubility across all compositions. Five distinct polymer weight ratios of P(VDF-TrFE):PLA were prepared to study the effect of composition on structural and functional properties: 0:100 (pure PLA), 25:75, 50:50, 75:25, 100:0 (pure P(VDF-TrFE)). 
After dissolution, a fixed amount (1 ml) of the mixed solutions was cast onto optically clean glass slides, which provided a smooth, non-reactive surface ensuring uniform film thickness and minimal interaction during solvent evaporation. The films were baked in a hot air oven at 140$^{\circ}$C for 2 hours, allowing complete solvent removal and promoting partial recrystallization, particularly beneficial for P(VDF-TrFE)'s $\beta$-phase development. The resulting films were carefully peeled off from the glass substrates and stored in desiccators to prevent moisture absorption prior to characterization (Fig.~\ref{fig1}).\\
\subsection{Characterization Tools used for Investigation}
\textbf{Thickness measurements:} For measuring thickness of the samples, Mitutoyo digital micrometre with $\pm$1$\mu$m accuracy is used. Thickness of each sample is measured at least 10 times and the average value is taken.\\
\textbf{X-Ray Diffraction (XRD):} Structural crystallinity and phase identification were examined using Bruker’s XRD with Cu K$\alpha$ radiation ($\lambda = 1.5406$~\text{\AA}), across a $2\theta$ range of \text{$10^\circ$- $35^\circ$} at room temperature.\\
\textbf{Fourier Transform Infrared (FTIR) spectroscopy:} FTIR (Bruker) spectrometer, within the 1900–600 cm$^{-1}$ range, is employed to identify functional group interactions and phase behavior. The Attenuated Total Reflectance (ATR) mode is used for better signal acquisition from thin films.\\ 
\textbf{Scanning Electron Microscopy:} Surface morphology and microstructural distribution of PLA and P(VDF-TrFE) phases are analysed using ZEISS field emission scanning electron microscope (FE-SEM). The dried films were sputter-coated with gold to enhance conductivity. Scanning Electron Microscopy (SEM) provided qualitative evidence of phase dispersion, surface smoothness, and morphological compatibility at the microscale.\\
\textbf{Differential Scanning Calorimetry (DSC)}: Thermal analysis was carried out using DSC (Netzsch) to study melting temperature (T$_m$), glass transition temperature (T$_g$), and crystallinity temperature (T$_c$). Approximately 5–10 mg of each film sample are sealed in aluminium pans and heated from 30$^{\circ}$C to 200$^{\circ}$C at 10$^{\circ}$C/min under nitrogen purge.\\ 
\textbf{Mechanical Properties:} Mechanical performance is assessed using a Universal Testing Machine. Films are cut into standardized specimens as per ISO-527-3, the tests are conducted at ambient temperature, 50\% relative humidity. The crosshead speed was 1 mm/min. The tensile strength, elongation at break, and modulus values are recorded. These properties are used to evaluate the impact of P(VDF-TrFE) content on the strength of the PLA matrix. A minimum of five replicates are tested per composition to ensure statistical significance.\\
\begin{figure*}
\centering
  \includegraphics[width=0.95\linewidth]{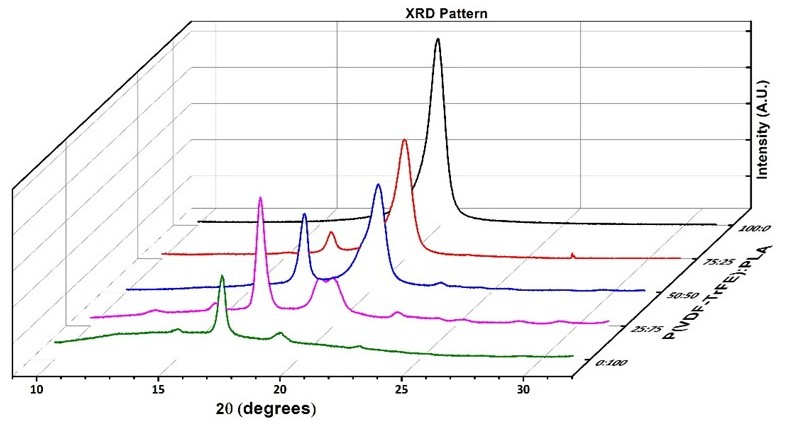}
  \caption{(a) Variation of P(VDF-TrFE):PLA blend composition, left bottle contains 0:100 (pure PLA), 25:75, 50:50, 75:25, and the right bottle contains 100:0 (pure P(VDF-TrFE)); (b) Casting technique adapted in film making.}
  \label{fig2}
\end{figure*}
\section{Results and Discussion}
\subsection{Phases and Microstructure}
Thickness of the films that are prepared from the varying blend composition are measured and their thicknesses are tabulated in Table~\ref{Table thickness}. Each film that casted is found to be nearly 40 $\mu$m in thickness\\
\begin{table}
\caption{Average thickness of blend films with varying blend composition.}
\centering
\begin{tabular}{c c}
\hline
Composition & Average Thickness\\
(P(VDF-TrFE):PLA) & ($\pm$ 1$\mu$ m)\\
\hline
 0:100 & 45\\
\hline
25:75 & 41\\
\hline
50:50 & 44\\
\hline
75:25 & 40\\
 \hline
 100:0 & 42\\
 \hline
\end{tabular}
\label{Table thickness}
\end{table}
\textbf{XRD Findings:} The XRD intensity within the diffraction angle range of 10$^{\circ}$ to 35$^{\circ}$ for various blend compositions is plotted and presented in Fig.~\ref{fig2}.  Generally, P(VDF-TrFE) typically exhibits strong diffraction peaks corresponding to its electroactive $\beta$-phase (e.g., at 20.6$^{\circ}$) \cite{R22}, whereas PLA presents peaks near 16.8$^{\circ}$ and 19.2$^{\circ}$ corresponding to (200)/(110) and (203) plane reflections respectively representing semicrystalline $\alpha$-form of PLA \cite{R23}. Possible variations in peak sharpness and position are examined to assess phase retention, potential co-crystallization, and the attenuation of peak intensity resulting from polymer blending. The XRD pattern of pure P(VDF-TrFE) (topmost black curve) exhibits a sharp and intense peak, around 20$^{\circ}$, which is the characteristic peak of the ferroelectric $\beta$-phase of P(VDF-TrFE). This strong peak corresponds to the (110)/(200) planes of the highly ordered $\beta$-phase, with the sharpness and intensity indicating a high degree of crystallinity \cite{R24}. Upon introducing 25\% PLA, a slight broadening of the P(VDF-TrFE) peaks is observed. The $\beta$-phase peaks at 20$^{\circ}$ shows a reduction in intensity, suggesting a partial disruption of crystalline ordering due to PLA incorporation. A minor peak appears at 16.9$^{\circ}$, it could be understood as the miscibility between the two polymers begins to affect the crystalline arrangements. The 50:50 (PVDF-TrFE:PLA) blend shows a further decrease in the intensity and sharpness of the $\beta$-phase peak of P(VDF-TrFE). Broader peaks with lower intensity dominate the spectrum, indicating a significant reduction in crystallinity. The humps appearing around 16.9$^{\circ}$ and 19.6$^{\circ}$, attributed to the presence of amorphous PLA, indicating the crystalline regions get affected due to the presence of P(VDF-TrFE). In the 25:75(PVDF-TrFE:PLA) blend, intensity of P(VDF-TrFE)-related crystalline peak become weak and diffuse, while the peak at 16.9$^{\circ}$ which corresponds to the PLA reaching its maximum intensity. The pattern is predominantly characterized by broad, low-intensity features of P(VDF-TrFE) and sharp reflecting peak of PLA, both amorphous and crystallinity behaviour convoluted from both the semi crystalline polymers. This observation suggests that the incorporation of P(VDF-TrFE) into the PLA matrix improves the nucleation and development of PLA’s crystallinity. The XRD profile of pure PLA (bottom most green curve) shows a broad amorphous halo centred between 12–20$^{\circ}$, confirming the predominantly amorphous nature of PLA. Pure PLA exhibited few characteristic peaks corresponding to the crystalline planes of (110/200), (203) and (015) planes at 16.8$^{\circ}$, 19.2$^{\circ}$ and 22.4$^{\circ}$ respectively, which coincide with the previous literature \cite{R25,R26}.\\
\begin{figure*}
\centering
  \includegraphics[width=0.92\linewidth]{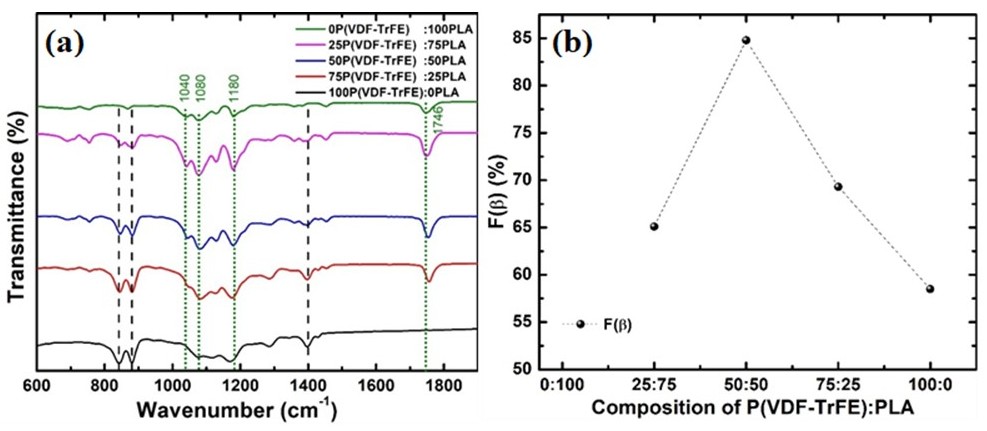}
  \caption{(a) FTIR spectrum of blend films with varying composition. Characteristic peaks of pure PLA are indicated green in color, while black in color for pure P(VDF-TrFE), (b) Degree of electroactive phase content F($\beta$) of P(VDF-TrFE) in the blend (dotted line is a visual guide)}
  \label{fig3}
\end{figure*}
\textbf{FTIR Investigation:} Fourier Transform Infrared spectroscopy (FTIR) data is used to establish the physical and chemical connections that exist between the constituent polymers in this experimental investigation. Transmittance values corresponding to the spectrum is plotted in the Fig.~\ref{fig3}(a). The characteristic peaks for -CF$_2$ (1170–1400 cm$^{-1}$) in P(VDF-TrFE) and the ester carbonyl (C=O, near $\sim$1750 cm$^{-1}$) in PLA are closely monitored. Shifts or intensity changes in these peaks were analysed to detect intermolecular interactions or physical blending effects. The peak corresponding to 868 cm$^{-1}$ represents the hydrogen bond in pure PLA. The noticeable peaks at 1040 cm$^{-1}$, 1080 cm$^{-1}$ and 1180 cm$^{-1}$ represent presence of C-O chemical bond in PLA \cite{R27}.  The peak at 1453 cm$^{-1}$ linked to the stretching of CH$_3$ bond while 1365 cm$^{-1}$ corresponds to the deformation of C-H bond. The 1746 cm$^{-1}$ peak represents stretching of -C=O bond of the ester group in pure PLA. An interesting observation is that the characteristic peak at 1746 cm$^{-1}$ gradually shifts to the higher wave number as the concentration of P(VDF-TrFE) increases and reaches a value of 1757 cm$^{-1}$ in 75P(VDF-TrFE):25PLA blend. It may suggest that the energy needed in stretching of -C=O is improved due to the addition of P(VDF-TrFE) \cite{R27}. This observation is in line with the DSC observation which would shortly be discussed in the later section. Since the amount of beta fraction F($\beta$) (ferroelectric content) is the exclusive property of P(VDF-TrFE), which indicates the crystalline arrangements of polymer chains. This value is calculated using a technique that is discussed in one of our previous works \cite{R10}. The amount of this electroactive content variation with composition variation is depicted in Fig.~\ref{fig3}(b). Its value initially increases, reaches a maximum at 50\% of P(VDF-TrFE) inclusion, then again, the value gradually decreases. Hence it could be understood that the blend ratio is playing a major role in polymer crystallization in P(VDF-TrFE) which is equally applicable in the case of PLA crystallization as well.\\
\begin{figure}
\centering
  \includegraphics[width=0.99\linewidth]{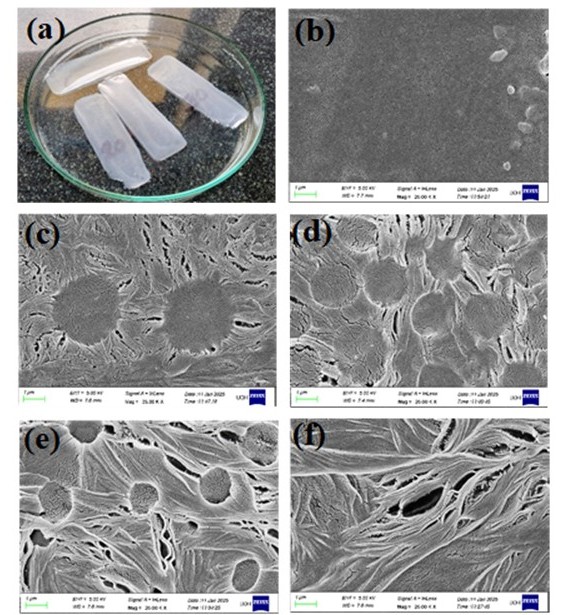}
  \caption{(a) Image of micron thick films prepared from the blends; SEM micrographs of (b) 0P(VDF-TrFE):100PLA, (c) 25P(VDF-TrFE):75PLA, (d) 50P(VDF-TrFE):50PLA, (e) 75P(VDF-TrFE):25PLA, (f) 100P(VDF-TrFE):0PLA samples, scale for the SEM images is uniform ($1~\mu\mathrm{m}$, as shown in imagery scale bar)}
  \label{fig4}
\end{figure}
\textbf{SEM Observations:} SEM images showed a distinct pattern in variation of surface microstructures of these films as shown in Fig.~\ref{fig4}. Differences in surface topology across the blend series are used to infer miscibility and interfacial behaviour.
The figure 4(a) shows the optical image of the specimen films. The SEM images imply a gradual variation of smooth to micro thin fibrillar structural transformation from Fig.~\ref{fig4}(b) to (f). Fig.~\ref{fig4} (b) shows a smooth, uniform surface with no distinct phase boundary topography for 100\% PLA. The observed microstructure could be attributed to the permeating and undisturbed crystalline and semicrystalline domains of the PLA matrix. From the Fig.~\ref{fig4}(c) inclusion of 25\% P(VDF-TrFE) in PLA matrix resulted relatively small islands of spherical structures of, phase separation, almost 3 microns in diameter be seen. These dense phase boundaries would influence the crystalline and amorphous packing in these semi crystalline polymers which could be reflected in electro active phase content and mechanical behaviour of the blend. The image Fig.~\ref{fig4}(d) shows a fine-grained uniform dispersion which may be an indication of optimal blend miscibility. This observation in line with the FTIR data that maximum $\beta$-phase is found in this composition. This could be interpreted as an Ideal phase mixing that promotes cooperative molecular interactions—both electroactive and mechanical properties expected to peak in the blend. Upon inclusion of more P(VDF-TrFE), Fig.~\ref{fig4}(e) depict the gradual decrement in the size of those spherical structures while an improvement in micro fibrillar dendritic structures. This may be an indication of dominance of P(VDF-TrFE) disrupting the ordering in packing which would be reflected as poor mechanical continuity and reduced functional synergy. Fig.~\ref{fig4}(f) shows the characteristic micro-fibrillar surface morphology of P(VDF-TrFE). Loss in structural reinforcement and electroactive alignments and fibrillar morphology may reflect pliability and loss of stiffness. From these images it is quite clear that despite having a common solvent, these two polymers show immiscible nature. SEM imaging supports the interplay between morphological ordering and functional performance, affirming the critical role of blend ratio in tailoring mechanical and electroactive behavior that supports the previous literature \cite{R10}.\\
\textbf{DSC Findings:} Differential Scanning Calorimetry (DSC) is used to investigate the physical and chemical changes that occur in a material during its thermal treatment. To eliminate the memory of thermal processing during synthesis of the samples, usually first melting endotherm is neglected, and the samples are subjected to subsequent controlled cooling for re-solidification. The second heating curves of the films of all the blends are investigated and the results are as plotted in Fig.~\ref{fig5}.
\begin{figure*}
\centering
  \includegraphics[width=0.95\linewidth]{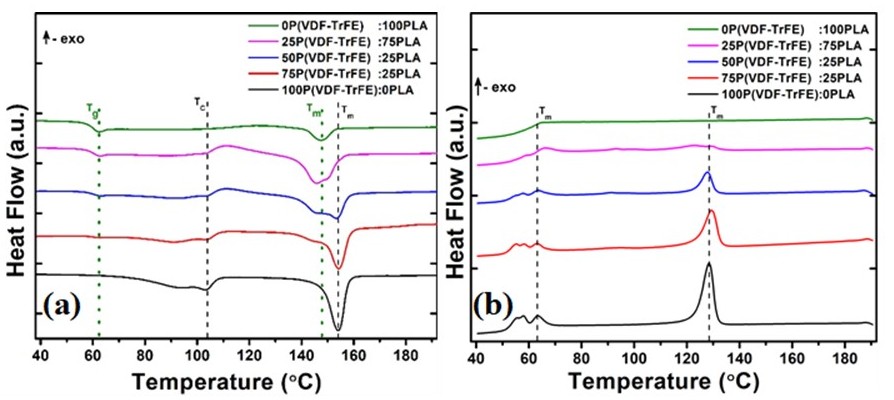}
  \caption{DSC thermograms for varying blend composition while (a) heating (b) cooling}
  \label{fig5}
\end{figure*}
The energy associated with a thermal transition can be understood by measuring the area under the curve of DSC measurement. The melting endotherm of P(VDF-TrFE) ($\sim$150–160$^{\circ}$C) and the glass transition of PLA ($\sim$60$^{\circ}$C) were used as thermal fingerprints. Shifts in these thermal events indicated possible phase interaction or restricted mobility in the blended states. From the endothermic transition peaks of Fig.~\ref{fig5}(a), it is noticed that melting of specimen 100P(VDF-TrFE):0PLA (means pure P(VDF-TrFE), (bottom most black curve)) has broadly two peaks (for easy identification, which are shown as black vertical dashed lines). The first broad peak on left side around 80 to 110$^{\circ}$C which is associated with ferro-electric to para-electric phase transition (T$_C$) while the right-side peak around 154$^{\circ}$C is crystal melting (T$_m$) of P(VDF-TrFE) \cite{R10}. Similarly, the endothermic transition of 0P(VDF-TrFE):100PLA (means pure PLA, (top most green curve)) has also two peaks (which are shown as green vertical dashed lines), the first left one around 62.5$^{\circ}$C, is due to the glass transition (T$_g$), while the right side one is due to the melting (T$_m$) of PLA which is near 148$^{\circ}$C. Interesting point to be noted here is, the crystallization temperature (T$_c$) and melting temperature (T$_m$) of PLA is falling in the ferro-para electric Curie transition (T$_C$) temperature window of P(VDF-TrFE). This thermal proximity could influence the co-crystallization behaviour of these semi-crystalline polymers one on other, that may lead to phase transitions of one polymer may influence the structural ordering or chain mobility of the other \cite{R28, R29, R30}. A gradual variation in melting curves can be seen from one pure polymer to the other. Interestingly, melting peaks of both the polymers are seen in the 50:50 composition blend, it indicates the independent thermal nature of these polymers in the blend. Peak broadening and slight depression with respect to the pure polymers is an indication of partial phase mixing and interaction-induced disorder \cite{R31}. The 25:75 blend shows the most well-defined melting behaviour of glass transition and crystal melting emanating from both the polymers which is an indication of good thermal stability of the blend. Enthalpic change during fusion ($\Delta$H$_f$) of these thermal transitions of the blends are tabulated in Table~\ref{table:enthalpy_detailed}. Intermediate blends have a gradual Enthalpy variation from pure PLA to pure P(VDF-TrFE).\\

\begin{table*}[htbp]
\caption{Enthalpy change ($\Delta H_f$) of blend films, obtained from DSC with varying blend composition.}
\centering
\begin{tabular}{c | c | c | c | c | c | c}
\hline
\textbf{(P(VDF-TrFE):PLA)} & $\Delta H_f$ (J/g) & $\Delta H_f$ (J/g) & $\Delta H_f$ (J/g) & $\Delta H_f$ (J/g) & $\Delta H_f$ (J/g) \\
& \textbf{$T_g$ of PLA} & \textbf{$T_c$ of PLA} & \textbf{$T_m$ of PLA} & \textbf{$T_c$ of P(VDF-TrFE)} & \textbf{$T_m$ of P(VDF-TrFE)} \\
\hline
0:100 & 2.59 & 4.63 & 6.38 & -- & -- \\
25:75 & 1.99 & 5.03 & 23.17 & 0.82 & 0.69 \\
50:50 & 0.72 & 10.64 & 18.00 & 7.16 & 5.65 \\
75:25 & 0.44 & 9.40 & 11.70 & 13.28 & 13.43 \\
100:0 & -- & -- & -- & 18.32 & 22.72 \\
\hline
\end{tabular}
\label{table:enthalpy_detailed}
\end{table*}

\begin{figure}
\centering
  \includegraphics[width=0.99\linewidth]{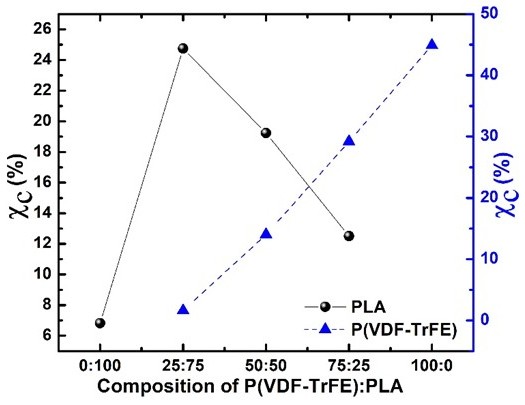}
  \caption{Degree of crystallinity with varying composition (the lines connecting the points are for visual guide)}
  \label{fig6}
\end{figure}

\begin{figure*}
\centering
  \includegraphics[width=0.99\linewidth]{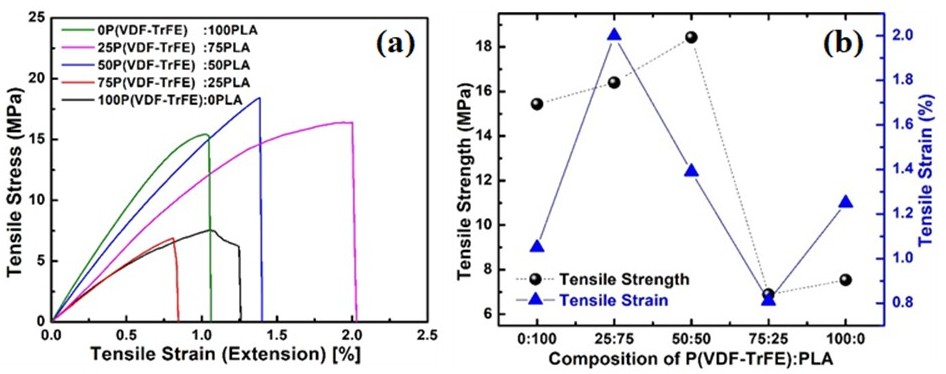}
  \caption{(a) stress-strain plots of blend films, (b) Variation of tensile strength on left and strain on right y- axis with varying composition on x- axis (lines connecting the symbols are guide to the eye))}
  \label{fig7}
\end{figure*}

In addition to this, exothermic thermal transitions of crystallisation peaks are shown in the Fig.~\ref{fig5}(b). For the pure P(VDF-TrFE) specimen (bottom most black curve) has ferroelectric crystallization transition between 50-70$^{\circ}$C and an amorphous to crystallization transition at 127$^{\circ}$C are clearly seen. For pure PLA (topmost green curve), glass transition can be seen at near 62$^{\circ}$C. A gradual variational transition can be seen from one pure polymer to the other polymer in between. The vertical black dashed lines show the temperature of melting for ferro electric and crystalline phases of pure P(VDF-TrFE) sample. The thermal properties such as glass transition (T$_g$), cold crystallization (T$_c$) and crystal melting temperatures (T$_m$) of PLA, as well as, ferro-electric to para-electric phase transition (T$_C$) and crystal melting (Tm) of P(VDF-TrFE) and corresponding to the intermediate blends are tabulated in Table~\ref{table:melting_crystallinity_simple}. Area under the endo/exo thermic peaks are deconvoluted to find the respective energy changes associated during thermal transition. \\
These values are used to find the degree of crystallinity ($\chi_C$) of PLA and P(VDF-TrFE) films using the equations,
\begin{equation}
    \chi_C=\frac{\Delta H\times 100}{\Delta H_{100\% crystalline}}
\end{equation}
where $\Delta H_{100\% crystalline}$ is the enthalpy change for 100\% crystalline P(VDF-TrFE) and 100\% crystalline PLA which is equal to 91.4 J/g and 93.6 J/g respectively \cite{R32, R33}. The values of degree of crystallinity of PLA and P(VDF-TrFE) measured from DSC, are plotted in Fig.~\ref{fig6}, it is interesting to highlight that the 25P(VDF-TrFE):75PLA composition has the highest degree of crystallinity. It could influence the mechanical behaviour of the films, which is discussed in the following section of mechanical investigation.\\
\begin{table*}[htbp]
\caption{Melting and crystallization temperatures, area deconvoluted energy (J/g), and degree of crystallinity of blend films with varying PLA and P(VDF-TrFE) composition.}
\centering
\begin{tabular}{c | c c c | c c | c c | c c}
\hline
\textbf{} & 
\multicolumn{3}{c|}{\textbf{PLA}} & 
\multicolumn{2}{c|}{\textbf{P(VDF-TrFE)}} & 
\multicolumn{2}{c|}{\textbf{Area Deconvoluted (J/g)}} & 
\multicolumn{2}{c}{\textbf{$\chi_C$ (\%)}} \\
\hline
(P(VDF-TrFE):PLA) & $T_g$ ($^{\circ}$C) & $T_c$ ($^{\circ}$C) & $T_m$ ($^{\circ}$C) & $T_c$ ($^{\circ}$C) & $T_m$ ($^{\circ}$C) & PLA & P(VDF-TrFe) & PLA & P(VDF-TrFE) \\
\hline
0:100 & 62.50 & 123.99 & 147.50 & -- & -- & 6.38 & -- & 6.8 & -- \\
25:75 & 62.50 & 111.99 & 145.99 & 91.49 & 149.49 & 23.17 & 1.51 & 24.7 & 1.6 \\
50:50 & 62.50 & 110.99 & 146.49 & 91.49 & 153.49 & 18.00 & 12.81 & 19.2 & 14.0 \\
75:25 & 62.50 & 112.99 & 144.99 & 91.49 & 153.99 & 11.70 & 26.71 & 12.5 & 29.2 \\
100:0 & -- & -- & -- & 94.49 & 153.99 & -- & 41.04 & -- & 44.9 \\
\hline
\end{tabular}
\label{table:melting_crystallinity_simple}
\end{table*}
\textbf{Mechanical Investigation:} The mechanical behaviour of P(VDF-TrFE):PLA blend films were evaluated via tensile testing using a Universal Testing Machine. The results of tensile
stress-strain profiles for different blend ratios are shown in Fig.~\ref{fig7}(a), and the associated mechanical parameters like, tensile stress at maximum load, strain at brake and tensile strain are summarized in Table~\ref{table:tensile}.\\
\textbf{Tensile Strength:} From the stress-strain curves, it is evident that the tensile strength of the blend films varies significantly with composition. Starting with the pure PLA film (0P(VDF-TrFE):100PLA), a tensile strength of 15.43 MPa and a strain at break of 2.58\% were observed, reflecting PLA’s inherent stiffness and moderate ductility. Upon incorporating 25\% P(VDF-TrFE) (25P(VDF-TrFE):75PLA), the tensile strength increased slightly to 16.40 MPa, along with a marginal improvement in elongation at brake to 2.93\%. The blend with equal proportions (50P(VDF-TrFE): 50PLA) exhibited the highest tensile strength of 18.43 MPa, indicating an optimal balance between the structural rigidity of PLA and the ductile, impact-modifying characteristics of P(VDF-TrFE). However, further increasing the P(VDF-TrFE) content led to a dramatic decline in tensile strength, with the 75P(VDF-TrFE):25PLA blend showing a significantly reduced value of 6.89 MPa and for the pure P(VDF-TrFE) film, it showed a tensile strength of 7.54 MPa. 
With the incorporation of P(VDF-TrFE) into PLA, the blend films demonstrated comparable or slightly enhanced elongation at break. The tensile strain initially increased, reached saturation, and subsequently declined—indicating improved flexibility in the blends but a reduced ability to bear mechanical load for higher load content. This mechanical trend aligns with prior studies on PVDF-based blends, where excessive incorporation of the soft phase results in diminished tensile properties due to decreased crystallinity and interfacial cohesion \cite{R34,R35}. These findings confirm that intermediate compositions, particularly the 25:75 and 50:50 blend compositions yield the most favourable combination of strength and toughness, which could be due to the effective interfacial phase interaction and stress distribution \cite{R36,R37}. This trend reflects the semi-crystalline structure of PLA, which contributes to higher tensile resistance. However, PLA's inherent brittleness is mitigated when blended with P(VDF-TrFE), a softer and more ductile copolymer. The 50:50 composition benefits from a synergistic balance of stiffness from PLA and flexibility from P(VDF-TrFE), which likely promotes enhanced load distribution and improved interfacial adhesion between phases \cite{R8,R17}.\\
\begin{table}[ht]
\caption{Tensile strength, strain at break and Tensile strain for varying blend composition films.}
\centering
\begin{tabular}{c c c c }
\hline
Composition & Tensile & Strain & Tensile \\ 
 & Strength & Break & Strength\\ 
 \hline
P(VDF-TrFE):PLA & (MPa) & (\%) & (\%) \\
\hline
0:100 & 15.43 & 2.58 & 1.05 \\
\hline
25:75 & 16.40 & 2.93 & 2.00 \\
\hline
50:50 & 18.43 & 2.88 & 1.39\\
\hline
75:25 & 6.89 & 2.90 & 0.81\\
\hline
100:0 & 7.54 & 2.13 & 1.25\\
\hline
\end{tabular}
\label{table:tensile}
\end{table}
\textbf{Tensile Strain:} The elongation behaviour of the P(VDF-TrFE):PLA blends, as indicated by the tensile strain, also exhibited a composition-dependent trend. The pure PLA film (0\% P(VDF-TrFE)) recorded a tensile strain of 1.05\%, which increased to the maximum in the 25\% P(VDF-TrFE) blend to 2.00\%. The 50:50 blend showed a comparable value of 1.39\%, while the 75:25 P(VDF-TrFE): PLA composition demonstrated the least tensile strain of 0.81\%, reflecting its least flexibility. Finally, for the 100\% P(VDF-TrFE) films, the tensile strain is found to be 1.25\%.
This observed trend highlights the plasticizing effect of P(VDF-TrFE) when incorporated into the PLA matrix in moderate amounts. The improved elongation at intermediate compositions may result from better chain mobility and interfacial interaction, which allow the material to undergo higher deformation before failure \cite{R38}. The peak elongation at the 25:75 composition suggests that smaller P(VDF-TrFE) content inclusion facilitates more segmental motion, contributing to flexibility \cite{R39}. However, upon more P(VDF-TrFE) inclusion, the decline in tensile strain is likely due to structural inhomogeneities or defects induced during solvent casting and thermal treatment, which limit its expected extensibility. These findings are in line with earlier studies, where improper crystallization or phase segregation in PVDF-rich systems negatively impacted mechanical resilience \cite{R34}. Thus, while P(VDF-TrFE) contributes to ductility, its optimal effect is realized only when blended with PLA in balanced proportions.\\
\begin{table*}[ht]
\caption{The morphology- structure- property observations of P(VDF-TrFE):PLA blends observed from
the above findings are summarized as below}
\centering
\begin{tabular}{c | c | c | c | c }
\hline
P(VDF-TrFE):PLA & Electroactive  & Morphology  & Crystallization in &Tensile Strength \\ 
 & Phase F($\beta$) (FTIR) & (SEM) & PLA (DSC) &  (MPa) \\
\hline
0:100 & Not detectable & Smooth,  & Low (6.8\%)  & 15.43\\
 & (Since Pure PLA) & continuous PLA matrix& – broad melting peak& \\
 \hline
25:75 & Moderate (65.1\%) & Dense and & Highest (24.7\%) & 16.40 \\
 & & ordered texture & -sharp peak & \\
\hline
50:50 & Maximum (84.8\%) & Uniform and & Moderate (19.2\%) & 18.43 \\
 & & fine dispersion & – broadened peak & \\
\hline
75:25 & Declining (69.3\%) & Irregular with & Low (12.5\%) & 6.89 \\
 & & phase boundaries& – suppressed peak & \\
\hline
100:0 & Reduced (58.5\%) & Soft, fibrillar morphology & Absent 
(no PLA phase) &  7.54\\
\hline
\end{tabular}
\label{table:enthalpy}
\end{table*}
\textbf{Effect of Blend Ratio:} The mechanical performance of the P(VDF-TrFE):PLA blends demonstrates a distinct dependence on the weight ratio of the two polymers, indicating significant phase interaction effects. At 0\% P(VDF-TrFE), the mechanical behaviour is dominated by the intrinsic stiffness and semicrystalline structure of PLA, yielding a moderate tensile strength (15.43 MPa) and limited flexibility (strain at break: 2.58\%). Introducing 25\% P(VDF-TrFE) into the PLA matrix results in a slight increase in both tensile strength (16.40 MPa) and strain (2.93\%), suggesting initial improvements in interfacial compatibility and load distribution \cite{R38}.
The 50:50 blend exhibits the most favourable mechanical response, achieving the highest tensile strength (18.43 MPa) and maintaining good ductility (2.88\%). This optimal composition likely benefits from synergistic interactions between the PLA and P(VDF-TrFE) phases. This observation in line with the Pipertzis et al observation which showed that phase dynamics and crystallization behaviour in P(VDF-TrFE) are sensitive to blend composition and external conditions. It supports the concept that intermediate compositions like 50:50 enable synergistic phase coupling, improving mechanical strength and ductility \cite{R40}. At this ratio, the ductile P(VDF-TrFE) domains may effectively hinder crack propagation, while the PLA phase contributes rigidity, resulting in better stress transfer and energy absorption. Such improvements in mechanical performance at balanced compositions have also been reported in PVDF-based binary systems, where miscibility or interpenetration of phases contributes to enhanced tensile behaviour \cite{R17}.
However, increasing the P(VDF-TrFE) content to 75\% and above leads to a significant deterioration in mechanical strength, dropping to 6.89 MPa at 75:25 and only 7.54 MPa in the pure P(VDF-TrFE) film. This reduction is attributed to poor dispersion and possible phase separation, which disrupts the stress transfer network across the film. Moreover, excessive P(VDF-TrFE) reduces the overall crystallinity and stiffness of the matrix, compromising its load-bearing capability. Similar observations were reported by Kong, Chang, and Wang, where high filler or secondary polymer content in PVDF-based composites resulted in non-uniform morphology and reduced tensile strength due to weak interfacial bonding \cite{R35}.
Overall, these findings confirm that mechanical optimization occurs only within a narrow compositional window. Excess P(VDF-TrFE) content results in imbalance between stiffness and ductility, leading to compromised performance. The 50:50 blend thus represents a critical point of mechanical synergy where both strength and elongation are maximized. The morphology- structure- property observations of P(VDF-TrFE):PLA blends observed from the above findings are summarized in Table~\ref{table:enthalpy}.
\section{Conclusion}
In this study, it is explored and reported for the first time, how electroactive phase content, morphology, crystallinity and mechanical properties interact within P(VDF-TrFE)-PLA blend films. Our findings reveal a clear correlation between these structural features and the resulting mechanical behavior, allowing for controlled tuning of strength and functionality. To enhance the mechanical properties of inherently brittle PLA, varying amounts of P(VDF-TrFE) were systematically incorporated, leading to the fabrication of nearly 40 $\mu$m thick free-standing films. Interestingly, the 25:75 and 50:50 polymer blends exhibit superior mechanical strength compared to each of their individual components. In the 25:75 (P(VDF-TrFE):PLA) composition, DSC analysis showed that PLA crystallized most efficiently. This is further supported by FTIR results, which indicated a significant presence of the electroactive $\beta$-phase, F($\beta$). The improved crystallization is likely driven by blend-induced cooperative molecular ordering, contributing to enhanced mechanical performance. Hence, this blend could find potential applications in the tactile sensors. At the 50:50 ratio, the blend demonstrated optimal interaction between piezoelectric and dielectric phases, along with effective phase mixing. This combination yielded a well-balanced profile of mechanical strength and electroactive content, making it highly suitable for applications such as sensors and 3D printing. With further increase in P(VDF-TrFE) content, the alignment of polymer chains might become less compatible. This led to a decline in PLA crystallinity and F($\beta$) content, resulting in softer, more flexible films. Despite their lower mechanical strength, these compositions are promising candidates for flexible electronics and wearable ferroelectrics due to their greater compliance.
\section{Acknowledgements}
The study was financially supported by funding from the seed grant of VNR Vignan Jyothi Institute of Engineering and Technology with grant number: VNRVJIET/H\&S/2021/01/seed.
\section{Conflicts of Interest}
The authors declare no conflicts of interest.
\section{Data Availability Statement}
The data that supports the findings of this study are available from the corresponding author(s) upon reasonable request.
%
%

%
\end{document}